# Spontaneous curvature in two-dimensional van der Waals heterostructures

Yuxiang Gao[1], Fenglin Deng[1], Ri He[2], Zhicheng Zhong[1,3]

[1] *Suzhou Institute for Advanced Research, University of Science and Technology of China, Suzhou 215123, China*

[2] *Key Laboratory of Magnetic Materials Devices & Zhejiang Province Key Laboratory of Magnetic Materials and Application Technology, Ningbo Institute of Materials Technology and Engineering, Chinese Academy of Sciences, Ningbo 315201, China*

[3] *School of Artificial Intelligence and Data Science, University of Science and Technology of China, Suzhou 215123, China*

**Abstract**

Two-dimensional (2D) van der Waals (vdW) heterostructures consist of different 2D crystals with diverse properties, constituting the cornerstone of the new generation of 2D electronic devices. Yet interfaces in heterostructures inevitably break bulk symmetry and structural continuity, resulting in delicate atomic rearrangements and novel electronic structures. In this paper, we predict that 2D interfaces undergo "spontaneous curvature", which means when two flat 2D layers approach each other, they inevitably experience out-of-plane curvature. Based on deep-learning-assisted large-scale molecular dynamics simulations, we observed significant out-of-plane displacements up to 3.8 Å in graphene/BN bilayers induced by curvature, producing a stable hexagonal moiré pattern, which agrees well with experimentally observations. Additionally, the out-of-plane flexibility of 2D crystals enables the propagation of curvature throughout the system, thereby influencing the mechanical properties of the heterostructure. These findings offer fundamental insights into the atomic structure in 2D vdW heterostructures and pave the way for their applications in devices.

**Introduction**

"The interface is the device", stated Nobel laureate Herbert Kroemer. He referred to the unique opportunities interfaces offer for creating tunable novel multifunctionalities, which are possible owing to the strong interaction among charge, spin, orbital, and structural degrees of freedom [1,2]. The quality of interfaces plays a dominant role in determining device performance, driving the development of interface engineering [3-5]. The atomically sharp interfaces are the quality goal and the eternal pursuit in the preparation of heterostructures at atomic resolution. However, interfaces often suffer from the significant strain effects and geometric defects, which result in the damage of atomically sharp interface [6-9]. Thanks to weaker interlayer van der Waals (vdW) interaction compared to intralayer chemical bonds in bulk interfaces, two-dimensional (2D) crystals with vastly different lattice constants can be easily stacked without experiencing large strain deformation or local disorders, enabling atomically sharp interfaces. Thus, 2D vdW heterostructures provide a platform that allows a far greater number of combinations than traditional bulk heterostructures [10-18].

Yet, due to the prevailing notion that 2D interfaces undergo minimal deformation, their atomic rearrangements are frequently neglected and simplified as rigid 2D layers in numerous studies[19,20]. Moreover, probing the atomic structure of 2D interfaces requires sub-atomic precision in experimental characterization and large-scale atomic simulations, making such investigations challenging. Recently, several studies have observed surface reconstruction at graphene/BN (Gr/BN) and transition metal dichalcogenide (TMD) bilayers, which is attributed to in-plane strain redistribution [21-27]. Nonetheless, 2D materials are more flexible out-of-plane than in-plane, so the out-of-plane displacement freedom and the curvature should be considered when exploring the atomic rearrangement at 2D vdW heterostructures. Here, we predict that 2D vdW heterostructures undergoes "spontaneous curvature" driven by the competition between stacking energy and deformation energy. The curvature additionally reduces the total energy of the entire system beyond the strain effect, playing a role similar to defects in 3D heterostructures.

**Results and discussions**

Stacking different 2D crystals inevitably leads to interfacial mismatch between layers, such as lattice mismatch, different crystallographic orientations or element types

[22,25,28-36]. Therefore, the stacking mode at 2D interfaces undergoes continuous changes, resulting in several energetically unfavorable stacking modes. Fig. 1b depicts the 1D atomic arrangement described by rigid model, surface reconstruction model and our spontaneous curvature model. The top 1D chain has a smaller lattice constant compared to the bottom 1D chain. The green color represents unstable stacking modes with high stacking energy, while the blue color represents stable stacking modes. If the atomic positions of 2D layers are fixed, as described by rigid model, the area for each stack mode is equivalent and unstable stacking modes occupy a significant portion. Apparently, this is not the optimal stacking distribution.

To decrease the proportion of unstable stacking structures, 2D interfaces undergo spontaneous atomic rearrangements. Surface reconstruction model was then proposed to describe the atomic rearrangements when considering the in-plane strain effect. The lattice with larger (smaller) lattice constant stretched (compressed) to quickly pass through the unstable stacking configuration. However, 2D materials always exhibit a high Young's modulus, thereby limiting their in-plane deformation [37]. Conversely, the bending modulus of 2D materials is nearly an order of magnitude smaller than their Young's modulus, potentially offering a more effective mechanism for adjusting atom positions [37-41].

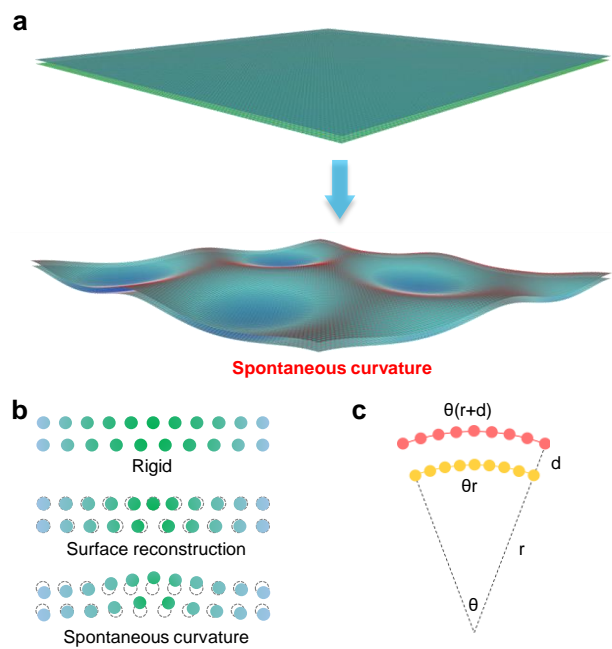

**Figure 1.** Spontaneous curvature model for 2D vdW heterostructure. (a) The atomic model of Gr/BN bilayer before and after relaxation, with the out-of-plane fluctuation amplitude amplified by a factor of 3. (b) 1D schematic comparing the rigid model, surface reconstruction model and spontaneous curvature model. The green color represents the unstable stacking state. (c) Geometry schematic of a 2D interface with curvature.

Fig. 1a shows the 3D initial flat model and the fully relaxed spontaneous curvature model based on a typical Gr/BN bilayer. The unrestricted stacking of pristine 2D monolayers leads to significant distortion in the out-of-plane direction, departing from their flat state. The periodic pattern aligns with the moiré pattern, with sharp bulge regions corresponding to domain walls and smooth concavities corresponding to domains. As shown in Fig. 1c, the curved 1D atomic chain exhibit a geometric structure similar to concentric circles. Thus, when the outer layer and inner layer have the same number of atoms, their equivalent lattice mismatch $\delta_c = d\kappa$, where $d$ represents the interlayer distance and $\kappa$ denotes the curvature. Thus, by adjusting the value of $\kappa$, the $\delta_c$ can be easily modulated.

Lattice mismatch intrinsically exists in 2D heterostructures and is the source of stacking modes variation. The curvature induced lattice mismatch $\delta_c$ plays a similar role on modulating stacking modes. If $\delta_c$ equals the original lattice mismatch between 2D lattices, the local stacking configuration can be retained without any bond stretching or compressing. If $\delta_c$ is opposite to the lattice mismatch between 2D lattices, wherein the outer layer has a smaller lattice constant and the inner layer has a larger lattice constant, the local stacking configuration changes more rapidly. Thus, curvature can modulate local stacking states independent of the in-plane strain. The curvature is limited by the balance between the gain in vdW interaction and the loss in bending energy. To our knowledge, the bending modulus of 2D materials is consistently smaller than their Young's modulus [37,42,43], thus the out-of-plane curvature provide a more effective mechanism to modulate local stacking configurations than in-plane strain.

Atomic rearrangements in 2D vdW heterostructures is primarily governed by the interplay between interlayer vdW interactions and intralayer lattice deformation. The total energy ($E$) of the relaxed system consists of two primary components: the vdW energy ($U$) and deformation energy ($\Pi$). The vdW energy and deformation energy for a given 2D heterostructure constructed by several 2D monolayers $\Omega_i$ are given by:

$$U[\psi] = \sum_j \sum_i \int \frac{1}{2} u_{ij}(\psi)\, d\Omega_i \tag{1}$$

$$\Pi[\Phi] = \sum_i \int \widehat{W}[\boldsymbol{C}_i(\Phi), \boldsymbol{K}_i(\Phi)]\, d\Omega_i \tag{2}$$

where $u$ is the vdW energy density (energy per unit undeformed area) functional and $\psi$ is the local stacking state. The $\Omega_i$ and $\Omega_j$ are neighboring 2D layers. $W$ is the stored deformation energy density functional determined by the left Cauchy-Green deformation tensor $\boldsymbol{C}_i$ and the Lagrangian curvature tensor $\boldsymbol{K}_i$ extracted from the deformation map $\Phi$ of the surface $\Omega_i$ [44]. The effective Lagrangian elasticity tensors for each 2D monolayer can be obtained as follows:

$$Y = 4 \frac{\partial^2 \widehat{W}}{\partial \boldsymbol{C}^2} \tag{3}$$

$$D = \frac{\partial^2 \widehat{W}}{\partial \boldsymbol{K}^2} \tag{4}$$

where Y is the in-plane stiffness and D is the out-of-plane stiffness of the surface, respectively.

Atomic rearrangements, including in-plane strain $\varepsilon$ and out-of-plane curvature $\kappa$, modulate the local lattice mismatch by $\delta = \delta_0 + \delta_\varepsilon + \delta_c$, where $\delta_0$ is the original lattice mismatch of the 2D layers without any relaxation, $\delta_\varepsilon = \boldsymbol{C}_i - \boldsymbol{C}_j$ represents the equivalent lattice mismatch change induced by in-plane strain, and $\delta_c = d\boldsymbol{K}$ denotes the equivalent lattice mismatch change induced by out-of-plane curvature. When considering a 2D interface with two monolayers possess isotropic in-plane stiffness and deformation maps, a simple relation between $\delta_\varepsilon$ and $\delta_c$ at the high-energy domain wall is as follows:

$$\frac{\delta_c}{\delta_\varepsilon} = f \frac{d^2 \overline{Y}}{4\overline{D}} \tag{5}$$

where $\overline{Y}$ and $\overline{D}$ denote the average in-plane and out-of-plane stiffness of 2D monolayers, respectively. $f$ represents a structure factor determined by deformation map. $d$ represents the interlayer spacing between the two monolayers (detailed derivation was provided in the Supplementary Material S2). According to this relation, the in-plane strain and out-of-plane curvature always co-exist. And the larger the $\overline{Y}/\overline{D}$, the more significant the role of curvature in atomic rearrangements. For typical 2D vdW heterostructures such as Gr/BN bilayers or TMD bilayers with larger Y compared to D, the impact of out-of-plane curvature on atomic rearrangements should be more significant than that of the in-plane strain.

To verify our hypotheses, molecular dynamics (MD) simulations with density functional theory (DFT) level accuracy assisted by deep learning were performed in the Gr/BN bilayers and $MoS_2/MoSe_2$ bilayers. Since Gr/BN and TMDs bilayers are the most intriguing 2D heterostructures with promising applications in next-generation electronic devices. In our simulations, the moiré supercell was commensurate to ensure that the original monolayers are flat and strain free. The lattice constant constants of Gr and BN monolayers are 2.468 Å and 2.512 Å, respectively. The original lattice mismatch of Gr and BN is approximately 1.8 %, thus a commensurate Gr/BN moiré superlattice should consist of a 56 × 56 Gr supercell and a 55 × 55 BN supercell with a periodic length of 14 nm. While after relaxation, we observed significant out-of-plane corrugation in Gr/BN bilayers. Detailed analyses of the atomic model and topological features of Gr/BN heterostructures will be discussed later. To confirm the universality of the curvature behavior, similar results of $MoS_2/MoSe_2$ were illustrated in Supplemental Material Fig. S2.

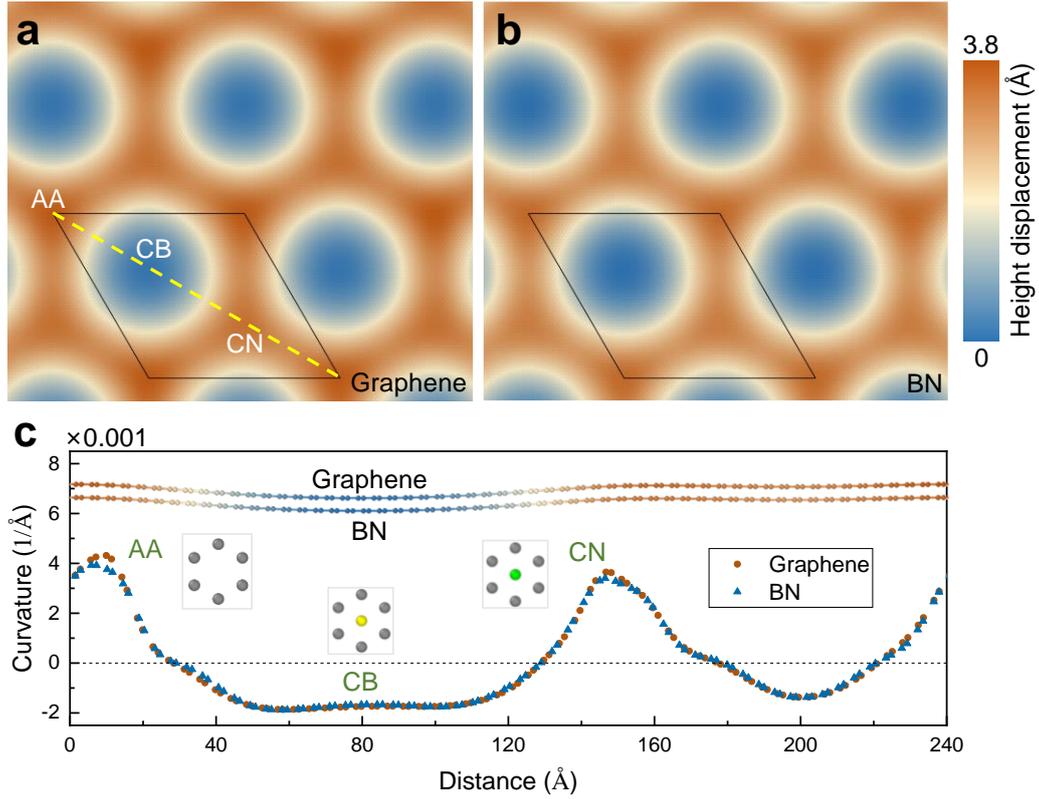

**Figure 2.** Spontaneous curvature in Gr/BN moiré superlattice. Height displacement distributions of (a) the Gr layer and (b) the BN layer. Atoms are colored according to their height relative to the lowest height in each layer, as shown in the right color bar. (c) The curvature calculated according to the atomic positions along the yellow dashed line in (a). The insets depict a side view of the yellow dashed line in (a) and the atomic structures of AA, CB and CN stacking modes. The gray, yellow and green balls represent C, N and B atoms, respectively.

Fig. 2a and 2b depict the out-of-plane topology of monolayer Gr and BN of a Gr/BN moiré superlattice with a periodic length of 138 Å. Height displacement refers to the difference in height between atoms within each layer and the lowest point of that layer. The height displacement distributions of the Gr and BN layers correspond to their moiré patterns. The height of the domain is significantly lower than that of the domain wall, with a maximum height difference of up to 3.8 Å. This height difference is an order of magnitude larger than the reported intrinsic ripple in Gr [45] or the differences in interlayer spacing among different stacking states of Gr/BN heterostructures [23]. At the lowest height, the Gr/BN layers exhibit CB stacking, while the stacking modes at the highest site and second highest site are AA and CN stacking, respectively. The height sequence is consistent with their stacking energy, with the stacking energy of AA stacking and CN stacking being 0.014 eV/atom and 0.013 eV/atom higher than that of CB stacking, respectively.

Fig. 2c illustrates the curvature calculated according to atomic positions of Gr and BN extracted along the yellow dashed line in Fig. 2a. The curvature of Gr and BN layers exhibits remarkable similarity. In the domain with CB stacking, the Gr and BN layers exhibit concavity, characterized by negative curvature of -0.002 Å$^{-1}$. The negative curvature in the CB stacking region persists over a considerable range of approximately 70 Å. Conversely, in the vertexes of the domain wall, the Gr and BN layers exhibit convexity, characterized by positive curvature of 0.004 Å$^{-1}$. According to our spontaneous curvature model, the equivalent lattice mismatch change $\delta_c$ at domain wall induced by out-of-plane curvature is calculated to be 0.014 with an interlayer distance of 3.48 Å. Meanwhile, the $\delta_\varepsilon$ induced by in-plane strain at domain wall is around 0.003. Thus, the impact of out-of-plane curvature on atomic rearrangements is nearly five times that of the in-plane strain in Gr/BN bilayers.

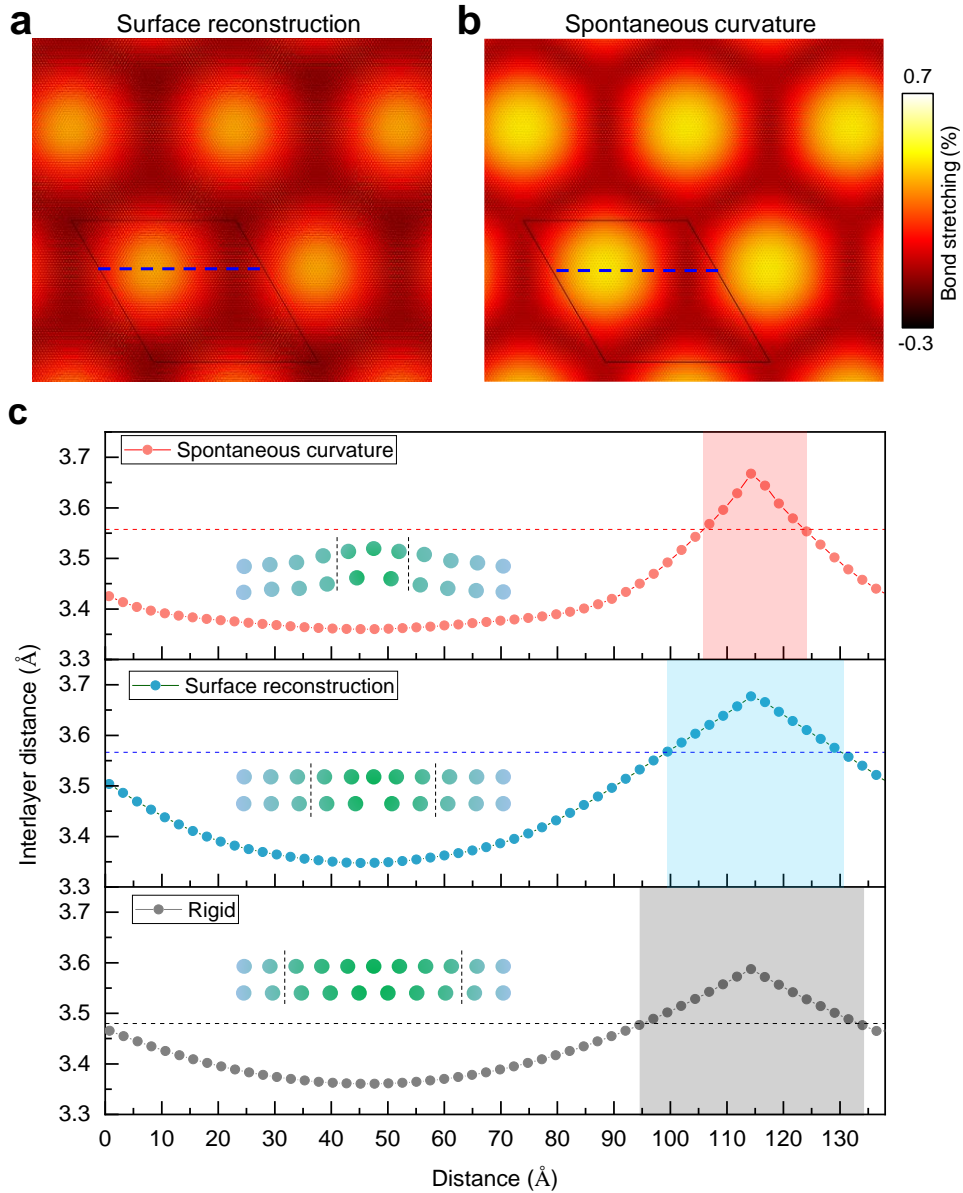

**Figure 3.** Atomic rearrangements of Gr/BN bilayers based on the surface reconstruction model and spontaneous curvature model. The bond length distribution of Gr layers in (a) the surface reconstruction model and (b) the spontaneous curvature model. Bond stretching represents the proportion by which the bond length exceeds the equilibrium bond length. (c) The interlayer distance of rigid model, surface reconstruction model and spontaneous curvature model.

According to our spontaneous curvature model, the opposing concavity directions result in a converse effect on modulating the local stacking configuration, expanding the domain while shrinking the domain wall. Fig. 3a and 3b indicate the bond length change of Gr in surface reconstruction model and spontaneous curvature model compared to the equilibrium bond length (1.424 Å). Previous experiment and theoretical studies both discovered bond stretching in domain and bond compressing in domain wall of Gr layers in Gr/BN moiré superlattices [23,46]. In spontaneous curvature model, the domain wall featured by bond compressing is notably narrower than that in surface reconstruction model. The results confirm our prediction that the curvature can further decrease the domain wall beyond in-plane strain effect.

To elucidate the impact of curvature on modulating local stacking states, we compared the variation in stacking states in Gr/BN moiré superlattice among the rigid model, surface reconstruction model, and spontaneous curvature model in Fig. 3e. The interlayer distance was calculated according to atomic arrangements extracted along the blue dashed lines in Fig. 3a, representing the distance between carbon atoms in Gr layer and their nearest atoms in BN layer. In the domain, the interlayer distance is minimal, while in the domain wall, the interlayer distance reaches its maximum. When the interlayer distance exceeds 97% of the maximum, it is considered part of the domain wall, as highlighted by colored shadows. The domain wall widths of rigid model, surface reconstruction model and spontaneous curvature model are 39 Å, 30 Å and 18 Å, respectively. Our spontaneous curvature model effectively reduces the domain wall width by 40% compared to surface reconstruction model. Furthermore, it should be noted that a plateau appears in the domain, indicating the maintenance of the most stable stacking configuration over a considerable range of 70 Å. The most stable stacking configuration is maintained with the range where concavity appears. Thus, the curvature of Gr and BN layers has significant effect on modulating their local stacking configurations.

Table 1. Total energies of the Gr/BN moiré superlattice.

| model | Rigid | Surface reconstruction | Spontaneous curvature |
|---|---|---|---|
| Energy (eV) | 0 | -10.50 | -14.95 |
| Domain wall width (Å) | 39 | 30 | 18 |

The expansion of low-energy domain and suppression of high-energy domain wall are expected to reduce the total energy of Gr/BN bilayers, as shown in Table 1. In comparison to the rigid model, the total energies of Gr/BN bilayers in surface reconstruction model and spontaneous curvature model are reduced by 10.50 eV and 14.95 eV, respectively. This suggests that curvature, in addition to in-plane strain in the surface reconstruction model, further decreases the energy of 2D heterostructures and stabilizes their structures. Because of the large energy gap, the curvature is robust under thermal fluctuation. Our MD simulations indicate that the curvature is promoted at finite temperature, as shown in Fig. 4. The curvature induced out-of-plane corrugation was maintained and the maximum of height displacement reaches 5.44 Å and 7.04 Å at 300 K and 600 K, respectively.

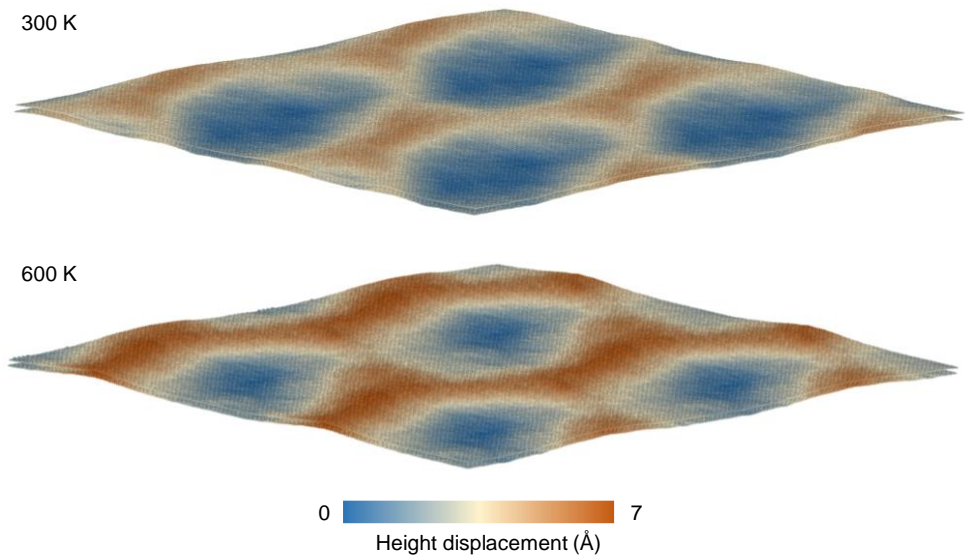

**Figure 4.** The height displacement distributions of Gr/BN bilayers at finite temperature.

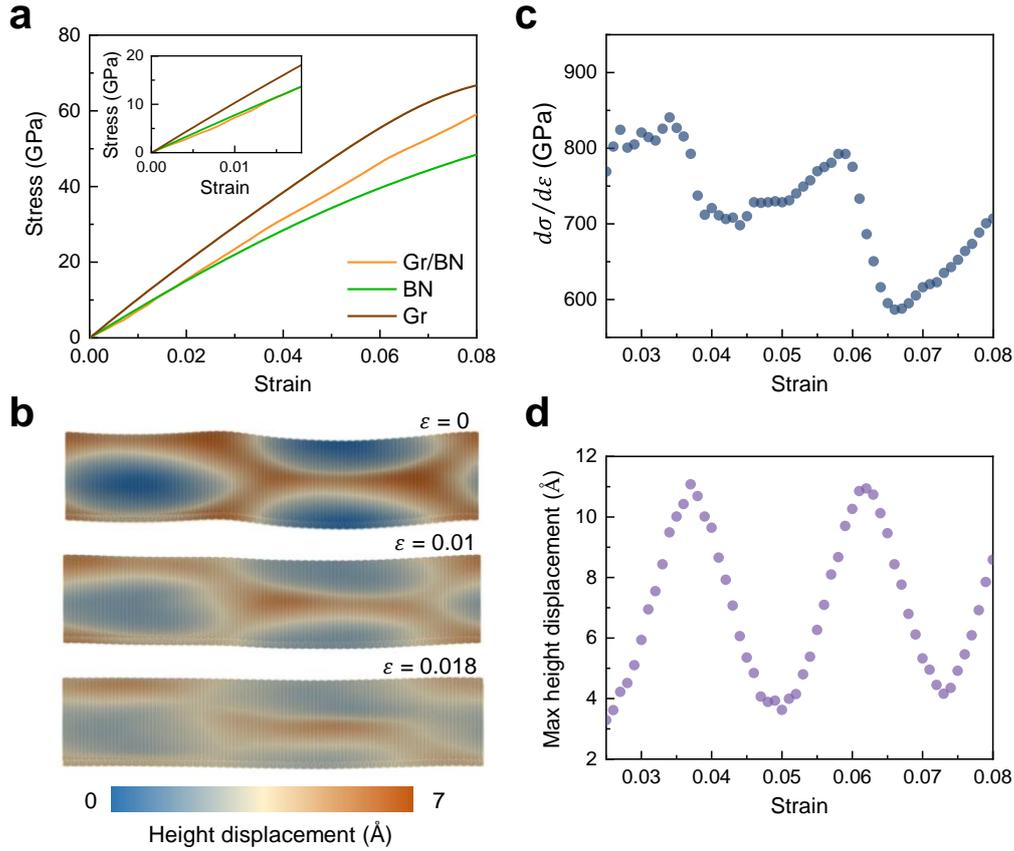

**Figure 5.** The curvature pattern and mechanical properties of Gr/BN heterostructures under external loading. (a) Stress-strain relations of Gr monolayer (Gr), BN monolayer (BN) and Gr/BN bilayers under uniaxial tensile loading along the *x* axis. Inset depicts the stress-strain relations of Gr, BN and Gr/BN under uniaxial tensile strains from 0 to 0.018. (b) The 3D atomic structures of Gr/BN bilayers under 0, 0.01 and 0.018 strain. Atoms are colored according to their height displacements as shown in the color bar below. (c) and (d) The first derivative of stress-strain curves ($d\sigma/d\varepsilon$) and max height displacement of Gr/BN bilayers under uniaxial loading along the *x* axis, respectively.

The curvature of Gr/BN bilayer is originated from the interplay between vdW interaction and lattice distortion, thus it is likely to be affected by external loading. To investigate the effect of external stress on the curvature behavior of the Gr/BN heterostructure, we applied non-equilibrium MD (NEMD) simulations at 1 K to minimize the effect of thermal fluctuations. A rectangular cell was used with x-axis along the armchair direction and y-axis along the zigzag direction. The model is 239 Å in length and 138 Å in width. By analyzing the atomic structure of Gr/BN bilayer, we found that as the strain increases, the out of plane corrugation gradually diminishes, as shown in Fig.5b. As strain reaches 0.018, the moiré pattern of Gr/BN bilayer is distorted and the domain wall network is broken, indicating that the coupling between the vdW interaction and lattice deformation is overpowered by the external loading. Additionally, we also investigate the curvature pattern of Gr/BN bilayer under biaxial strain. As shown in in Supplemental Material Fig. S3, as the strain increases, the out-of-plane displacement decreases significantly. Thus, the curvature pattern of Gr/BN bilayer is sensitive to the external strain and can be readily modulated by small external loadings.

The vanishing of the out-of-plane moiré pattern affects the mechanical properties of the Gr/BN bilayer. We compared stress–strain curves of Gr monolayer, BN monolayer and Gr/BN heterostructure under uniaxial tensile loading, as presented in Fig. 5a. In principle, the strength of Gr/BN bilayer should be in the middle of that of Gr and BN monolayers. However, when the strain is less than 0.018, the stress of the Gr/BN bilayer is lower than that of both the Gr and BN monolayers, indicating that the Gr/BN bilayers were softened by the out-of-plane moiré pattern. As strain exceeds 0.018 and the out-of-plane moiré pattern is totally destroyed, the strength of Gr/BN bilayer recovers.

Besides, the stress-strain relation of the Gr/BN system exhibits nonlinear behavior with fluctuations as the strain exceeds 0.018. Fig. 5c depicts the $d\sigma/d\varepsilon$ curve under uniaxial tensile strain ranging from 0.025 to 0.08 with periodic fluctuations. The $d\sigma/d\varepsilon$ reflects the resistance of Gr/BN bilayer to external loadings. To explore the source of periodic fluctuations, we analyzed the atomic structure of the Gr/BN bilayer. MD simulations found that the Gr/BN heterostructure curves at the direction perpendicular to the applied uniaxial tensile loading. The out-of-plane topology of the Gr/BN heterostructure transforms into a stripe pattern parallel to the uniaxial tensile loading. We use max height displacement to evaluate the out-of-plane corrugation of the Gr/BN bilayer. As shown in Fig. 4d, a periodic variation of max height displacement with increasing strain was observed and the period is corresponding to that of the $d\sigma/d\varepsilon$ curve. The $d\sigma/d\varepsilon$ curve undergoes a steep descent at peaks of the max height displacement, indicating the competition and conversion between strain energy as well as bending energy and confirming the effects of curvature pattern on the mechanical properties of the entire system.

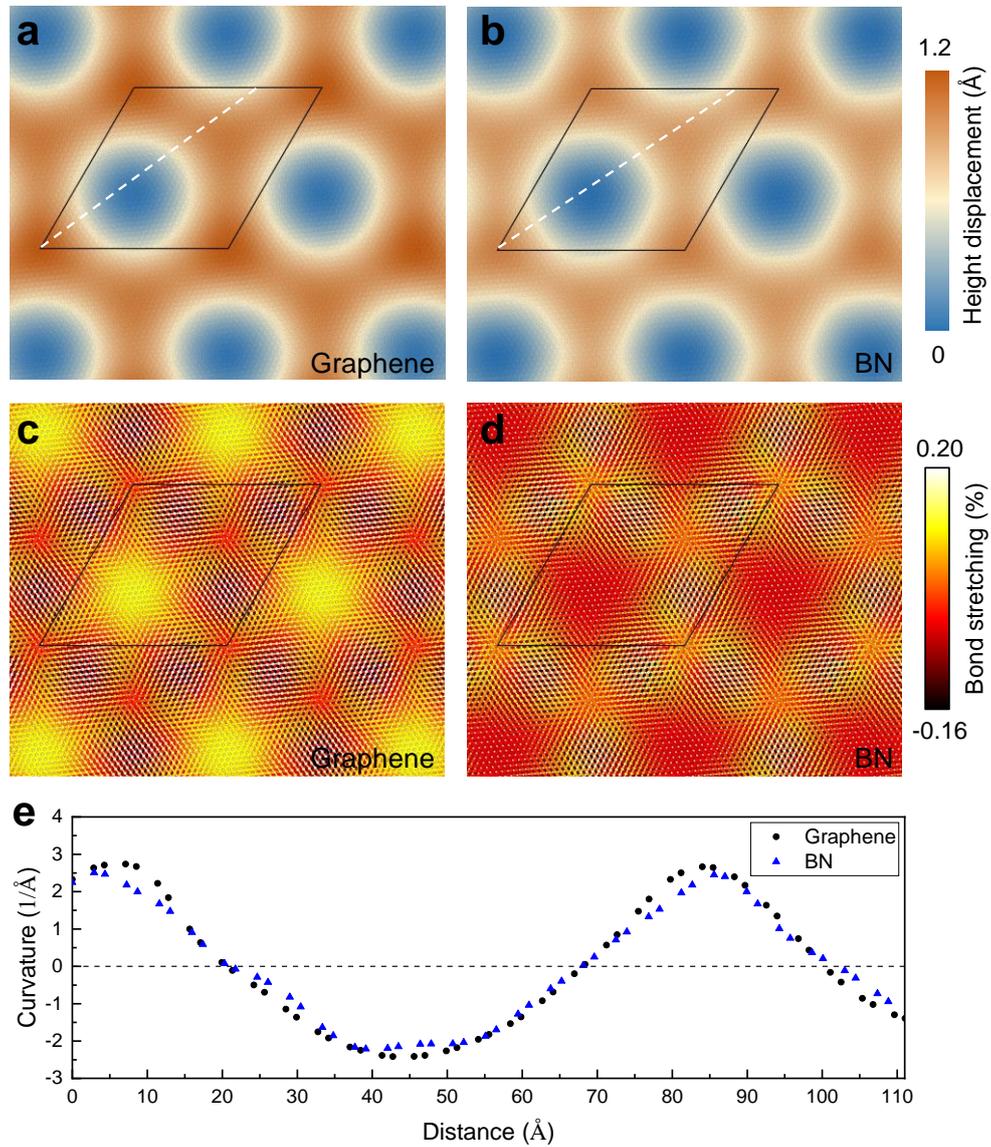

**Figure 6.** The geometry deformation in twisted Gr/BN heterostructures. The out-of-plane topology of (a) the Gr layer and (b) the BN layer. The bond length distribution of (c) the Gr layer and (d) the BN layer. (e) The curvature of Gr and BN layers calculated according to the atomic positions along the yellow dashed line in (a) and (b), respectively.

We also investigate the effect of lattice orientation on the curvature of Gr/BN bilayer. Fig. 6a and 6b depict the height displacement distribution of Gr and BN layers with a twist angle φ ~ 1.5°. The periodic length of the moiré superlattice is 79 Å, significantly smaller than that of the Gr/BN moiré superlattice with practically aligned crystallographic axes (φ = 0°). Consequently, the maximum height displacement of the twisted Gr/BN bilayer also decreased to 1.2 Å. Additionally, the distribution of stacking states in the twisted Gr/BN bilayer is consistent with that of the aligned Gr/BN heterostructure. At the lowest height in the twisted Gr/BN heterostructure, the stacking state is CB stacking, while at the highest and second-highest sites, AA and CB stacking are observed, respectively.

Fig. 6c and 6d indicate the bond length distribution of Gr and BN layers in twisted Gr/BN bilayer. The bond length is stretched within the domain and compressed within the domain wall in the Gr layer. However, there is no sharp change in bond length across the domain wall in each moiré unit cell, making it difficult to identify the edge of the domain wall. The distinct difference in bond length distribution between the Gr/BN heterostructure with φ ~ 1.5° and φ ~ 0° was previously observed and recognized as the commensurate-incommensurate transition [46]. Based on the explicit atomic arrangement from MD simulations, we found that the bond length is highly anisotropic within the twisted Gr/BN moiré superlattice and exhibits in-plane rotation [47,48]. Local bond extension and suppression co-exist in the transition region where the stacking states changes from AA to CN stacking.

While the bond length distribution undergoes significant changes with varying lattice orientations of Gr and BN layers, their curvature behaviors remain similar to the perfectly aligned Gr/BN bilayers. Fig. 6e depicts the curvature of [1$\bar{1}$0] lattice orientation in Gr and BN layer, calculated using atomic positions extracted along the white dashed lines in Fig. 6a and 6b, respectively. In both layers, the minimal curvature is approximately -0.002 Å$^{-1}$ at the domain, while the maximal curvature is around 0.003

Å$^{-1}$ at the domain wall. These results elucidate that the curvature behavior is robust despite the interlayer mismatch of lattice orientations.

**Conclusions and perspectives**

In summary, we proposed a spontaneous curvature model to elucidate the atomic rearrangements in 2D vdW heterostructures. The model was confirmed through MD simulations with DFT level accuracy assisted by deep-learning neural networks in Gr/BN and TMD bilayers. Upon relaxation, Gr/BN bilayers exhibit smooth concavity with negative curvature in the low-energy domain and sharp convexity with positive curvature in the high-energy domain wall. Such out-of-plane curvature additionally narrowed the domain wall width and sufficiently reduced the total energy of the entire system beyond the in-plane strain effect. The curvature is robust despite the thermal fluctuations and interlayer mismatch of lattice orientations. Furthermore, we found that the curvature is sensitive to strain conditions and can be readily modulated by small external loadings. The 3D pattern induced by curvature affects the mechanical properties of the Gr/BN bilayer. When the moiré pattern exists, the Gr/BN bilayer is softened compared to Gr and BN monolayers. As the moiré pattern vanishes and transforms to a strip pattern, the strength of Gr/BN bilayers recovers and presents periodic fluctuations.

Based on our MD simulations, curvature potentially offers the unique opportunity to modulate the mechanical properties of vdW heterostructures by controlling the atomic rearrangement at 2D interfaces. Previous studies have found that the 3D strain field creates a pseudo-magnetic field in excess of 300 tesla in monolayer Gr with large lattice distortion [49]. Thus, the electronic properties of 2D materials can also be profoundly altered by controlling their atomic rearrangements [50]. Inspired by these observations, we consider that this exploration can be extended to other properties such as spintronics, ferroelectricity, magnetism, and thermodynamics, which may be affected by atomic rearrangements in 2D interfaces, potentially offering intriguing possibilities to some novel phenomena.

**Acknowledgements**

This work was supported by the National Key R&D Program of China (Grants No. 2021YFA0718900 and No. 2022YFA1403000), Key Research Program of Frontier Sciences of CAS (Grant No. ZDBS-LY-SLH008), National Nature Science Foundation of China (Grants No. 12304049 and No. 12204496) and the Zhejiang Provincial Natural Science Foundation (Grants No. Q23A040003).

**Contributions**

Z. Z. conceived the project and supervised the research. Y. G. performed the theoretical analysis, deep potential training and molecular dynamics simulations. Y. G. organized and wrote the paper. F. D., and R. H. helped to revise the paper and provide scientific discussion when this study encountered problems.

**Competing interests**

The authors declare no competing interests.

**Data availability**

All the training datasets, DP model, and configurations of Gr/BN and TMD heterostructures to reproduce the results contained in this paper are available in AIS Square website:

https://www.aissquare.com/datasets/detail?pageType=datasets&name=Gr_BN_bilayer&id=271